\documentclass[aps,prl,twocolumn,superscriptaddress,preprintnumbers,amsmath,amssymb]{revtex4}
\usepackage{graphicx} % Include figure files
\usepackage{dcolumn}  % Align table columns on decimal point
\usepackage{rotating}
\usepackage{booktabs}
\usepackage{SIunits} % for unit
\usepackage{graphicx}
\usepackage{dcolumn}
\usepackage{diagbox}
\usepackage{bm}
\usepackage{color}
\usepackage[english]{babel}
\usepackage{titlesec}
\usepackage{multirow}
\usepackage[colorlinks,linkcolor=blue,urlcolor=blue,anchorcolor=green,citecolor=blue,breaklinks=true]{hyperref}
\usepackage{epstopdf}
\usepackage{enumitem}
\usepackage{hhline}
\usepackage{array}

\usepackage{threeparttable}

\newcommand{\BR}{{\cal B}}

\newcommand{\pip}{\pi^+}

\newcommand{\EE}{e^+e^-}
\begin{document}
\hyphenpenalty=10000
\tolerance=1000
\vspace*{-3\baselineskip}
%\resizebox{!}{3cm}{\includegraphics{belle.eps}}

%\input{pub608}

%%% Paper:    Omega_c to Omega l nu
%%% Journal:  Physical Review Letters
%%% Contacts: Y.B. Li (liyb@pku.edu.cn)
%%%           C.P. Shen (shencp@fudan.edu.cn)
%%% Non-responding authors or those who said NO are commented out.
%%% ====================================================================
%%% Click the RELOAD button on your web browser to see the updated file.
%%% ====================================================================
%%% Use \input{author} to insert this material into your latex file.
%%%%% Force institutions to appear in alphabetical order when typeset.
\noaffiliation
\affiliation{Department of Physics, University of the Basque Country UPV/EHU, 48080 Bilbao}
\affiliation{University of Bonn, 53115 Bonn}
\affiliation{Brookhaven National Laboratory, Upton, New York 11973}
\affiliation{Budker Institute of Nuclear Physics SB RAS, Novosibirsk 630090}
\affiliation{Faculty of Mathematics and Physics, Charles University, 121 16 Prague}
%%%\affiliation{Chiba University, Chiba 263-8522}
\affiliation{Chonnam National University, Gwangju 61186}
\affiliation{Chung-Ang University, Seoul 06974}
\affiliation{University of Cincinnati, Cincinnati, Ohio 45221}
\affiliation{Deutsches Elektronen--Synchrotron, 22607 Hamburg}
%%%\affiliation{Duke University, Durham, North Carolina 27708}
\affiliation{Institute of Theoretical and Applied Research (ITAR), Duy Tan University, Hanoi 100000}
\affiliation{University of Florida, Gainesville, Florida 32611}
\affiliation{Department of Physics, Fu Jen Catholic University, Taipei 24205}
\affiliation{Key Laboratory of Nuclear Physics and Ion-beam Application (MOE) and Institute of Modern Physics, Fudan University, Shanghai 200443}
%%%\affiliation{Justus-Liebig-Universit\"at Gie\ss{}en, 35392 Gie\ss{}en}
\affiliation{Gifu University, Gifu 501-1193}
%%%\affiliation{II. Physikalisches Institut, Georg-August-Universit\"at G\"ottingen, 37073 G\"ottingen}
\affiliation{SOKENDAI (The Graduate University for Advanced Studies), Hayama 240-0193}
\affiliation{Gyeongsang National University, Jinju 52828}
\affiliation{Department of Physics and Institute of Natural Sciences, Hanyang University, Seoul 04763}
\affiliation{University of Hawaii, Honolulu, Hawaii 96822}
\affiliation{High Energy Accelerator Research Organization (KEK), Tsukuba 305-0801}
\affiliation{J-PARC Branch, KEK Theory Center, High Energy Accelerator Research Organization (KEK), Tsukuba 305-0801}
\affiliation{National Research University Higher School of Economics, Moscow 101000}
\affiliation{Forschungszentrum J\"{u}lich, 52425 J\"{u}lich}
%%%\affiliation{Hiroshima University, Higashi-Hiroshima, Hiroshima 739-8530}
\affiliation{IKERBASQUE, Basque Foundation for Science, 48013 Bilbao}
\affiliation{Indian Institute of Science Education and Research Mohali, SAS Nagar, 140306}
\affiliation{Indian Institute of Technology Bhubaneswar, Satya Nagar 751007}
\affiliation{Indian Institute of Technology Guwahati, Assam 781039}
\affiliation{Indian Institute of Technology Hyderabad, Telangana 502285}
\affiliation{Indian Institute of Technology Madras, Chennai 600036}
\affiliation{Indiana University, Bloomington, Indiana 47408}
\affiliation{Institute of High Energy Physics, Chinese Academy of Sciences, Beijing 100049}
\affiliation{Institute of High Energy Physics, Vienna 1050}
\affiliation{Institute for High Energy Physics, Protvino 142281}
%%%\affiliation{Institute of Mathematical Sciences, Chennai 600113}
\affiliation{INFN - Sezione di Napoli, I-80126 Napoli}
\affiliation{INFN - Sezione di Roma Tre, I-00146 Roma}
\affiliation{INFN - Sezione di Torino, I-10125 Torino}
\affiliation{Iowa State University, Ames, Iowa 50011}
\affiliation{Advanced Science Research Center, Japan Atomic Energy Agency, Naka 319-1195}
\affiliation{J. Stefan Institute, 1000 Ljubljana}
%%%\affiliation{Kanagawa University, Yokohama 221-8686}
\affiliation{Institut f\"ur Experimentelle Teilchenphysik, Karlsruher Institut f\"ur Technologie, 76131 Karlsruhe}
\affiliation{Kavli Institute for the Physics and Mathematics of the Universe (WPI), University of Tokyo, Kashiwa 277-8583}
%%%\affiliation{King Abdulaziz City for Science and Technology, Riyadh 11442}
\affiliation{Department of Physics, Faculty of Science, King Abdulaziz University, Jeddah 21589}
\affiliation{Kitasato University, Sagamihara 252-0373}
\affiliation{Korea Institute of Science and Technology Information, Daejeon 34141}
\affiliation{Korea University, Seoul 02841}
%%%\affiliation{Kyoto Sangyo University, Kyoto 603-8555}
\affiliation{Kyungpook National University, Daegu 41566}
%%%\affiliation{Universit\'{e} Paris-Saclay, CNRS/IN2P3, IJCLab, 91405 Orsay}
%%%\affiliation{\'Ecole Polytechnique F\'ed\'erale de Lausanne (EPFL), Lausanne 1015}
\affiliation{P.N. Lebedev Physical Institute of the Russian Academy of Sciences, Moscow 119991}
%%%\affiliation{Liaoning Normal University, Dalian 116029}
\affiliation{Faculty of Mathematics and Physics, University of Ljubljana, 1000 Ljubljana}
\affiliation{Ludwig Maximilians University, 80539 Munich}
\affiliation{Luther College, Decorah, Iowa 52101}
\affiliation{Malaviya National Institute of Technology Jaipur, Jaipur 302017}
%%%\affiliation{University of Malaya, 50603 Kuala Lumpur}
\affiliation{Faculty of Chemistry and Chemical Engineering, University of Maribor, 2000 Maribor}
\affiliation{Max-Planck-Institut f\"ur Physik, 80805 M\"unchen}
\affiliation{School of Physics, University of Melbourne, Victoria 3010}
\affiliation{University of Mississippi, University, Mississippi 38677}
%%%\affiliation{University of Miyazaki, Miyazaki 889-2192}
\affiliation{Moscow Physical Engineering Institute, Moscow 115409}
\affiliation{Graduate School of Science, Nagoya University, Nagoya 464-8602}
%%%\affiliation{Kobayashi-Maskawa Institute, Nagoya University, Nagoya 464-8602}
\affiliation{Universit\`{a} di Napoli Federico II, I-80126 Napoli}
\affiliation{Nara Women's University, Nara 630-8506}
\affiliation{National Central University, Chung-li 32054}
%%%\affiliation{National United University, Miao Li 36003}
\affiliation{Department of Physics, National Taiwan University, Taipei 10617}
\affiliation{H. Niewodniczanski Institute of Nuclear Physics, Krakow 31-342}
\affiliation{Nippon Dental University, Niigata 951-8580}
\affiliation{Niigata University, Niigata 950-2181}
%%%\affiliation{University of Nova Gorica, 5000 Nova Gorica}
\affiliation{Novosibirsk State University, Novosibirsk 630090}
%%%\affiliation{Okinawa Institute of Science and Technology, Okinawa 904-0495}
\affiliation{Osaka City University, Osaka 558-8585}
\affiliation{Pacific Northwest National Laboratory, Richland, Washington 99352}
\affiliation{Panjab University, Chandigarh 160014}
%%%\affiliation{Peking University, Beijing 100871}
\affiliation{University of Pittsburgh, Pittsburgh, Pennsylvania 15260}
\affiliation{Punjab Agricultural University, Ludhiana 141004}
%%%\affiliation{Research Center for Electron Photon Science, Tohoku University, Sendai 980-8578}
\affiliation{Research Center for Nuclear Physics, Osaka University, Osaka 567-0047}
\affiliation{Meson Science Laboratory, Cluster for Pioneering Research, RIKEN, Saitama 351-0198}
%%%\affiliation{Theoretical Research Division, Nishina Center, RIKEN, Saitama 351-0198}
%%%\affiliation{RIKEN BNL Research Center, Upton, New York 11973}
\affiliation{Dipartimento di Matematica e Fisica, Universit\`{a} di Roma Tre, I-00146 Roma}
\affiliation{Department of Modern Physics and State Key Laboratory of Particle Detection and Electronics, University of Science and Technology of China, Hefei 230026}
%%%\affiliation{Seoul National University, Seoul 08826}
\affiliation{Showa Pharmaceutical University, Tokyo 194-8543}
\affiliation{Soochow University, Suzhou 215006}
\affiliation{Soongsil University, Seoul 06978}
%%%\affiliation{University of South Carolina, Columbia, South Carolina 29208}
%%%\affiliation{Stefan Meyer Institute for Subatomic Physics, Vienna 1090}
\affiliation{Sungkyunkwan University, Suwon 16419}
\affiliation{School of Physics, University of Sydney, New South Wales 2006}
\affiliation{Department of Physics, Faculty of Science, University of Tabuk, Tabuk 71451}
\affiliation{Tata Institute of Fundamental Research, Mumbai 400005}
%%%\affiliation{Excellence Cluster Universe, Technische Universit\"at M\"unchen, 85748 Garching}
\affiliation{Department of Physics, Technische Universit\"at M\"unchen, 85748 Garching}
\affiliation{School of Physics and Astronomy, Tel Aviv University, Tel Aviv 69978}
\affiliation{Toho University, Funabashi 274-8510}
%%%\affiliation{Department of Physics, Tohoku University, Sendai 980-8578}
\affiliation{Earthquake Research Institute, University of Tokyo, Tokyo 113-0032}
\affiliation{Department of Physics, University of Tokyo, Tokyo 113-0033}
\affiliation{Tokyo Institute of Technology, Tokyo 152-8550}
\affiliation{Tokyo Metropolitan University, Tokyo 192-0397}
%%%\affiliation{Utkal University, Bhubaneswar 751004}
\affiliation{Virginia Polytechnic Institute and State University, Blacksburg, Virginia 24061}
\affiliation{Wayne State University, Detroit, Michigan 48202}
\affiliation{Yamagata University, Yamagata 990-8560}
\affiliation{Yonsei University, Seoul 03722}
  \author{Y.~B.~Li}\affiliation{Key Laboratory of Nuclear Physics and Ion-beam Application (MOE) and Institute of Modern Physics, Fudan University, Shanghai 200443} % Fudan
  \author{C.~P.~Shen}\affiliation{Key Laboratory of Nuclear Physics and Ion-beam Application (MOE) and Institute of Modern Physics, Fudan University, Shanghai 200443} % Fudan
  \author{I.~Adachi}\affiliation{High Energy Accelerator Research Organization (KEK), Tsukuba 305-0801}\affiliation{SOKENDAI (The Graduate University for Advanced Studies), Hayama 240-0193} % KEK
% \author{K.~Adamczyk}\affiliation{H. Niewodniczanski Institute of Nuclear Physics, Krakow 31-342} % Krakow
% \author{J.~K.~Ahn}\affiliation{Korea University, Seoul 02841} % Korea
  \author{H.~Aihara}\affiliation{Department of Physics, University of Tokyo, Tokyo 113-0033} % Tokyo
  \author{S.~Al~Said}\affiliation{Department of Physics, Faculty of Science, University of Tabuk, Tabuk 71451}\affiliation{Department of Physics, Faculty of Science, King Abdulaziz University, Jeddah 21589} % Tabuk
  \author{D.~M.~Asner}\affiliation{Brookhaven National Laboratory, Upton, New York 11973} % BNL
  \author{H.~Atmacan}\affiliation{University of Cincinnati, Cincinnati, Ohio 45221} % Cincinnati
% \author{V.~Aulchenko}\affiliation{Budker Institute of Nuclear Physics SB RAS, Novosibirsk 630090}\affiliation{Novosibirsk State University, Novosibirsk 630090} % BINP
  \author{T.~Aushev}\affiliation{National Research University Higher School of Economics, Moscow 101000} % HSE
  \author{R.~Ayad}\affiliation{Department of Physics, Faculty of Science, University of Tabuk, Tabuk 71451} % Tabuk
% \author{T.~Aziz}\affiliation{Tata Institute of Fundamental Research, Mumbai 400005} % Tata
  \author{V.~Babu}\affiliation{Deutsches Elektronen--Synchrotron, 22607 Hamburg} % DESY
% ...
% Lebedev
% \author{V.~Zhulanov}\affiliation{Budker Institute of Nuclear Physics SB RAS, Novosibirsk 630090}\affiliation{Novosibirsk State University, Novosibirsk 630090} % BINP
\collaboration{The Belle Collaboration}

%\preprint{\vbox{ \hbox{   }
		%\hbox{Belle DRAFT {\it 17-07}}
		%\hbox{Intended for {\it P.R.L}}
		%\hbox{Author: Y. B. Li, C. P. Shen}
		%\hbox{Committee: Christoph SCHWANDA (chair),}
		%\hbox{ $~~~~~~~~~~~~~~~~$Martin BESSNER, Mizuki SUMIHAMA}
%		\hbox{Belle Preprint \# 2021-29}
%		\hbox{KEK Preprint \# 2021-34}
%	}
%}

\title{\boldmath First test of Lepton Flavor Universality in the charmed baryon decays $\Omega^{0}_{c} \to \Omega^{-} \ell^{+} \nu_{\ell}$ using data of the Belle experiment}
%\boldmath First observation of $\Omega_{c}^{0} \to \Omega^{-} \mu^{+} \nu_{\mu}$ and measurements of the branching fractions of semileptonic decays $\Omega_{c}^{0} \to \Omega^{-} \ell^{+} \nu_{\ell}$}

%%%% >>>>> insert the authorlist here. BEFORE the abstract !!!!! <<<<<
%%%% >>>>> from the authorship confirmation web page
%%% Name the file author.tex and use \input{author} to insert into your latex file.
%\author{Author}\affiliation{affiliation}
%\collaboration{The Belle Collaboration}
%\noaffiliation
%% end author list

%%% Paper:   Omega0  SL decay
%%% Journal:  Physical Review Letters
%%% Contacts: Y.B. Li (liyb@fudan.edu.cn)
%%%           C.P. Shen (shencp@fudan.edu.cn)
%%% Non-responding authors or those who said NO are commented out.
%%% ====================================================================
%%% Click the RELOAD button on your web browser to see the updated file.
%%% ====================================================================
%%% Use \input{author} to insert this material into your latex file.
%%%%% Force institutions to appear in alphabetical order when typeset.

\noaffiliation

\begin{abstract}

We present the first observation of the $\Omega_{c}^{0} \to \Omega^{-} \mu^{+} \nu_{\mu}$  decay and present measurements of the branching fraction ratios of the $\Omega_{c}^{0} \to \Omega^{-} \ell^{+} \nu_{\ell}$ decays
compared to the reference mode $\Omega_{c}^{0} \to \Omega^{-} \pi^+$, ($\ell = e$ or $\mu$). This analysis is based on 89.5 fb$^{-1}$, 711 fb$^{-1}$, and 121.1 fb$^{-1}$ data samples collected with the Belle detector at the KEKB asymmetric-energy $e^+e^-$
collider at the center-of-mass energies of 10.52 GeV, 10.58 GeV, and 10.86 GeV, respectively.
The $\Omega_{c}^{0}$ signal yields are extracted by fitting $M_{\Omega \ell}$ and $M_{\Omega\pi}$ spectra.
The branching fraction ratios ${\cal B}(\Omega_{c}^{0} \to \Omega^{-} e^{+} \nu_{e})/{\cal B}(\Omega_{c}^{0} \to \Omega^{-} \pi^+)$ and
${\cal B}(\Omega_{c}^{0} \to \Omega^{-} \mu^{+} \nu_{\mu})/{\cal B}(\Omega_{c}^{0} \to \Omega^{-} \pi^+)$ are measured to be $1.98 \pm0.13~({\rm stat.}) \pm 0.08~({\rm syst.})$ and $ 1.94 \pm 0.18~({\rm stat.}) \pm 0.10~({\rm syst.})$, respectively.
The ratio of ${\cal B}(\Omega_{c}^{0} \to \Omega^{-} e^{+} \nu_{e})/{\cal B}(\Omega_{c}^{0} \to \Omega^{-} \mu^{+} \nu_{\mu})$
is measured to be $1.02 \pm 0.10~({\rm stat.}) \pm 0.02~({\rm syst.})$, which is consistent with the expectation of lepton flavor universality.

\end{abstract}

\maketitle

%%%% >>>> keep the final version single-spaced
\tighten

{\renewcommand{\thefootnote}{\fnsymbol{footnote}}}
\setcounter{footnote}{0}

%%%%%%%%%%%%%%%%%%%%%%%%%%%%%%%%%%%%%%
%%%%%%%%%   introduction  %%%%%%%%%%%%
%%%%%%%%%%%%%%%%%%%%%%%%%%%%%%%%%%%%%%
%\linenumbers

In the Standard Model (SM), the charged weak current interaction has an identical
coupling to all lepton generations, known as lepton flavor universality (LFU). However, experiments have found tantalizing deviations from LFU in $b \to c\ell \nu_{\ell}$ and $b \to s\ell\ell$ decays~\cite{RD_sum,RK_exp1,RK_exp2,RD_exp1,RD_exp2,RD_exp3}, especially an evidence of LFU breaking with a 3.1 standard deviations on branching fraction ratio $\BR(B^{+}\to K^{+} \mu^{+}\mu^{-})/\BR(B^{+}\to K^{+} e^{+} e^{-})$ at the LHCb experiment~\cite{LU_LHCb}.
Since a violation of LFU is a clear sign of new physics~\cite{LFU_theory1,LFU_theory2,LFU_theory3,LFU_theory4,LFU_theory5},
tests of LFU in additional semileptonic decays of heavy quarks are well motivated.

Lying in the transition region between the perturbative and non-perturbative energy scales of quantum chromodynamics (QCD), charmed baryons play an important role in studies of strong and weak interactions, especially via the investigations of their semileptonic decays~\cite{slRMP,HQS,HQS2}. Their decay amplitudes are the product of a well-understood leptonic current describing the lepton system and a more complicated hadronic current for the quark transition, which helps to measure SM parameters such as CKM matrix elements and study the details of decay dynamics.

Due to the low production rates and/or high background levels of current experiments, the study of charmed baryon decays is statistically limited. Thus far, semileptonic decays of $\Lambda_{c}^{+}$ and $\Xi^{0}_c$ have only been partially studied, and LFU is found to be conserved within uncertainties~\cite{lcbes1,lcbes2,BES_lc_Inc,belle_Xic0}. The sole result on semileptonic decays of $\Omega_c^0$ is CLEO's observation of $11.4 \pm 3.8$ events of $\Omega_{c}^{0} \to \Omega^{-} e^{+} \nu_{e}$, with a branching fraction ratio of $\BR(\Omega_{c}^{0} \to \Omega^{-} e^{+} \nu_{e})/\BR(\Omega_{c}^{0} \to \Omega^{-} \pi^{+})$ measured to be $2.4\pm1.2$~\cite{cleo_omgc}. Compared with the $\frac{1}{2}^{+} \to \frac{1}{2}^{+} $ transitions $\Lambda^{+}_{c} \to \Lambda^{0} $ and $\Xi^{0}_{c} \to \Xi^{-}$, the $\frac{1}{2}^{+} \to \frac{3}{2}^{+}$ decay $\Omega_{c}^{0} \to \Omega^{-} $ contains two more form factors in the hadronic current, which makes it more difficult to predict the decay rate theoretically~\cite{epjcOmgc_sl}. The predicted branching fraction $\BR(\Omega_{c}^{0} \to \Omega^{-} \ell^{+} \nu_{\ell})$ varies between 0.005 and 0.127 in light-front quark models~\cite{epjcOmgc_sl,qian}, heavy quark expansion~\cite{plbOmgc_sl}, and quark models~\cite{prcOmgc_sl}. Although the theoretical predictions on $\Omega_{c}^{0}$ semileptonic decay widths differ by more than an order of magnitude, the ratios between the $e$ and $\mu$ modes are stable and can be compared with the current experimental measurement to test LFU.

We note that the lifetime of $\Omega_{c}^{0}$ has been recently updated from $(69 \pm 12) \times 10^{-15}$ s~\cite{PDGold} to $(268 \pm 26) \times 10^{-15}~{\rm s}$~\cite{LHCb_Omegac_LF,PDG}. A precise study of the $\Omega_{c}^{0}$ is crucial to test the theoretical models as well as understand the $\Omega_c^{0}$ lifetime by comparing the measured branching fractions and corresponding theoretical predictions~\cite{HQE1,HQE2,HQE3,HQE4,CBcirca2021}, especially for its semileptonic decay since constructive interference between the $s$ quarks can result in a large semileptonic decay width~\cite{plbOmgc_sl,epjcOmgc_sl_abnor}.

In this Letter, we present a study of the semileptonic decays of $\Omega_{c}^{0} \to \Omega^{-} \ell^{+} \nu_{\ell}$ using data samples of 89.5 fb$^{-1}$, 711 fb$^{-1}$, and 121.1 fb$^{-1}$ collected by the Belle detector
at the KEKB asymmetric-energy collider~\cite{KEKB} at the center-of-mass energies of 10.52 GeV, 10.58 GeV, and 10.86 GeV, respectively, which is 66 times larger than the data set used in CLEO's analysis~\cite{cleo_omgc}. Inclusion of charge-conjugate states is implicit unless otherwise stated in this analysis. $\Omega_{c}^{0}$ are produced in the process $\EE \to c\bar{c}  \to \Omega_{c}^{0} + anything$, while $\Omega^{-}$ baryons are reconstructed via the $\Lambda K^-$ mode, where $\Lambda$ decays into $p\pi^{-}$. Branching fraction ratios of $\Omega_{c}^{0} \to \Omega^{-} \ell^{+} \nu_{\ell}$ to the reference mode $\Omega_{c}^{0} \to \Omega^{-} \pi^{+}$ are measured. The precision of $\BR(\Omega_{c}^{0} \to \Omega^{-} e^{+} \nu_{e})/\BR(\Omega_{c}^{0} \to \Omega^{-} \pip)$ is significantly improved compared to the previous result~\cite{cleo_omgc}. The previously unobserved $\Omega_c^0 \to \Omega^-\mu^+\nu_\mu$ decay	is also studied. LFU is thus probed in the decays $\Omega_{c}^{0}\to\Omega^{-}\ell^{+}\nu_{\ell}$ for the first time.

%%%%%%%%%%%%%%%%%%%%%%%%%%%%%%%%
%%   BELLE Detector
%%%%%%%%%%%%%%%%%%%%%%%%%%%%%%
The Belle detector is a large-solid-angle magnetic spectrometer that
consists of a silicon vertex detector, a 50-layer central drift chamber,
an array of aerogel threshold Cherenkov counters, a barrel-like arrangement
of time-of-flight scintillation counters, and an electromagnetic calorimeter comprised of CsI(Tl) crystals;
all these components are located inside a superconducting solenoid coil that provides a 1.5 T magnetic field. An iron flux-return located outside of the coil is instrumented	to detect $K^{0}_{L}$ mesons and identify muons (KLM). The direction of the $e^{+}$ momentum is defined as the $z$-axis direction.
The detector is described in detail elsewhere~\cite{Belle}.

%%%%%%%%%%%%%%%%%%%%%%%%%%%%%%%%%%%%%%%%%
%%%%%% event selection  %%%%%%%%%%%%%%%%%
%%%%%%%%%%%%%%%%%%%%%%%%%%%%%%%%%%%%%%%%%
To optimize the signal selection criteria and calculate the signal reconstruction efficiency, we use Monte Carlo (MC) simulated events.
The $\EE \to c\bar{c}$ process, and the signal $\Omega_c^0$ semileptonic decays are simulated with the {\sc pythia} with matrix element model~\cite{pythia}.
The $\Omega_{c}^{0} \to \Omega^{-} \pi^{+}$ decay is generated with {\sc EvtGen}~\cite{evtgen}.
The simulated $\Upsilon(4S)\to B \bar{B}$, $\Upsilon(5S) \to B_s^{(*)}\bar{B}_s^{(*)}$, $\Upsilon(5S) \to B^{(*)}\bar{B}^{(*)}(\pi)$, and $\Upsilon(5S)\to\Upsilon(4S)\gamma$ events with $B=B^+$ or $B^0$,
and $e^+e^- \to q\bar{q}$ events with $q=u,~d,~s$ or $~c$ at the center-of-mass energies of data
are used as background samples after removing the signal events.
The MC events are processed with a detector simulation based on {\sc geant3}~\cite{geant3}. The background sources and fit methods described later are also validated with simulated generic samples~\cite{zhouxy_topo}.

Except for the charged tracks from $\Omega^-$ decays,
the impact parameters perpendicular to and along the $e^{+}$ beam direction
are required to be less than 0.5 cm and 4.0 cm, respectively,
and the transverse momentum in the lab frame must be higher than 0.1 GeV/$c$.
For charged tracks, information from different detector subsystems is
combined to form the likelihood $\mathcal{L}_{i}$ for species $i$, where $i= e,~\mu,~\pi$,~$K$, or $p$~\cite{pid}.
A track with a likelihood ratio $\mathcal{L}_K/(\mathcal{L}_K + \mathcal{L}_\pi)> 0.6$ is
identified as a kaon, while a track with $\mathcal{L}_K/(\mathcal{L}_K + \mathcal{L}_\pi)<0.4$ is
treated as a pion~\cite{pid}. With this selection, the kaon (pion)
identification efficiency is about 94\% (98\%), while 2\% (5\%) of
the pions (kaons) are misidentified as kaons (pions).
A track with a likelihood ratio $\mathcal{L}_e/(\mathcal{L}_e+\mathcal{L}_{{\rm non}-e}) > 0.9$
is identified as an electron~\cite{lke}.
The $\gamma$ conversions are suppressed by
examining all combinations of an $e^{\pm}$ track with other oppositely-charged tracks
in the event that are identified as $e^{\mp}$,
and requiring an $\EE$ invariant mass larger than $0.4$ GeV/$c^{2}$.
Tracks with $\mathcal{L}_{\mu}/(\mathcal{L}_{\mu}+\mathcal{L}_{K}
+\mathcal{L}_{\pi})>0.9$ are considered as muon candidates~\cite{lku}.
The muon tracks also should hit at least five layers of the KLM subdetector,
and can not be identified as kaons by requiring $\mathcal{L}_{K}/(\mathcal{L}_K + \mathcal{L}_\pi)< 0.4$ to suppress backgrounds with kaons. With the above selections,
the efficiencies of electron and muon identifications are 98\% and 76\%, respectively, with the pion fake rates less than 2\%.

The $\Lambda$ baryons are reconstructed in the decay $\Lambda \to p \pi^-$
and selected if $|M_{p \pi^-}-m_{\Lambda}|<3.5$ MeV/$c^2$ (about three times the invariant mass resolution ($\sigma$)). Here and throughout the text,
$M_i$ represents a measured invariant mass and $m_i$ denotes the
nominal mass of the particle $i$~\cite{PDG}.
The proton track from $\Lambda$ decay is required to satisfy $\mathcal{L}_{p}/(\mathcal{L}_{\pi} + \mathcal{L}_p)> 0.2$ and $\mathcal{L}_{p}/(\mathcal{L}_{K} + \mathcal{L}_p)> 0.2$. These requirements identify protons with an efficiency of 95\% and the contamination from pions and kaons is less than 1\%. We define the $\Omega^-$
signal region as $|M_{\Lambda K^-}-m_{\Omega^-}|<3.5$ MeV/$c^{2}$ ($\sim$$3\sigma$). Since the background components of the $M_{\Lambda K}$ distributions can be described by a horizontal straight line, the $\Omega^-$ mass sidebands are chosen as 13 MeV/$c^2$ $<|M_{\Lambda K^-} - m_{\Omega^{-}}| < $ 27 MeV/$c^2$, which is four times the width of the signal region for facilitating the normalization in the following fits.
To suppress the combinatorial background,
we require the flight directions of $\Lambda$ and $\Omega^{-}$ candidates,
which are reconstructed from their fitted production and decay vertices,
to be within five degrees of their momentum directions in both 3D space and the plane perpendicular to the $z$-axis in the lab frame.
\begin{figure*}[htbp]
	\begin{center}
		\includegraphics[width=5cm]{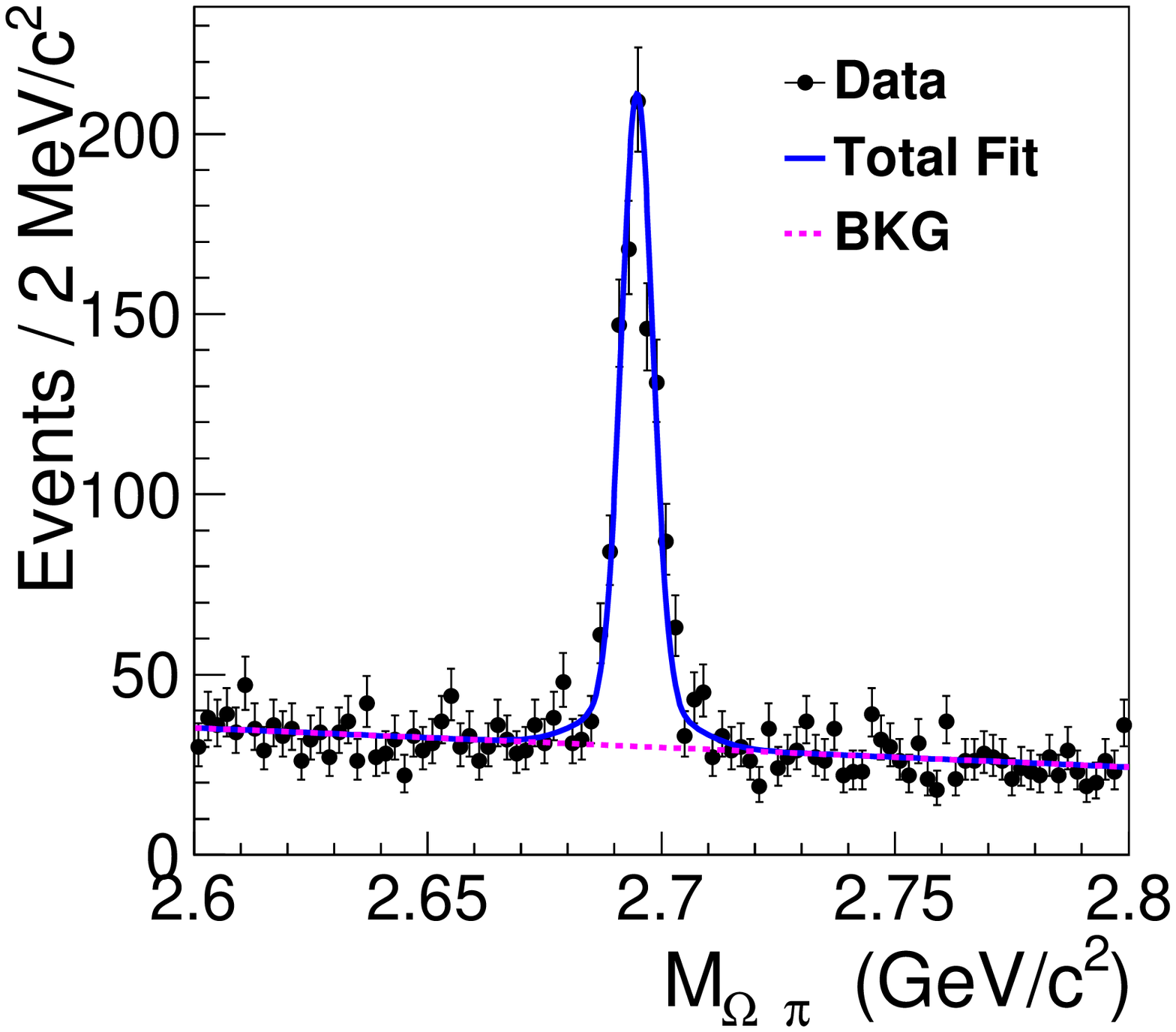}
		\includegraphics[width=5cm]{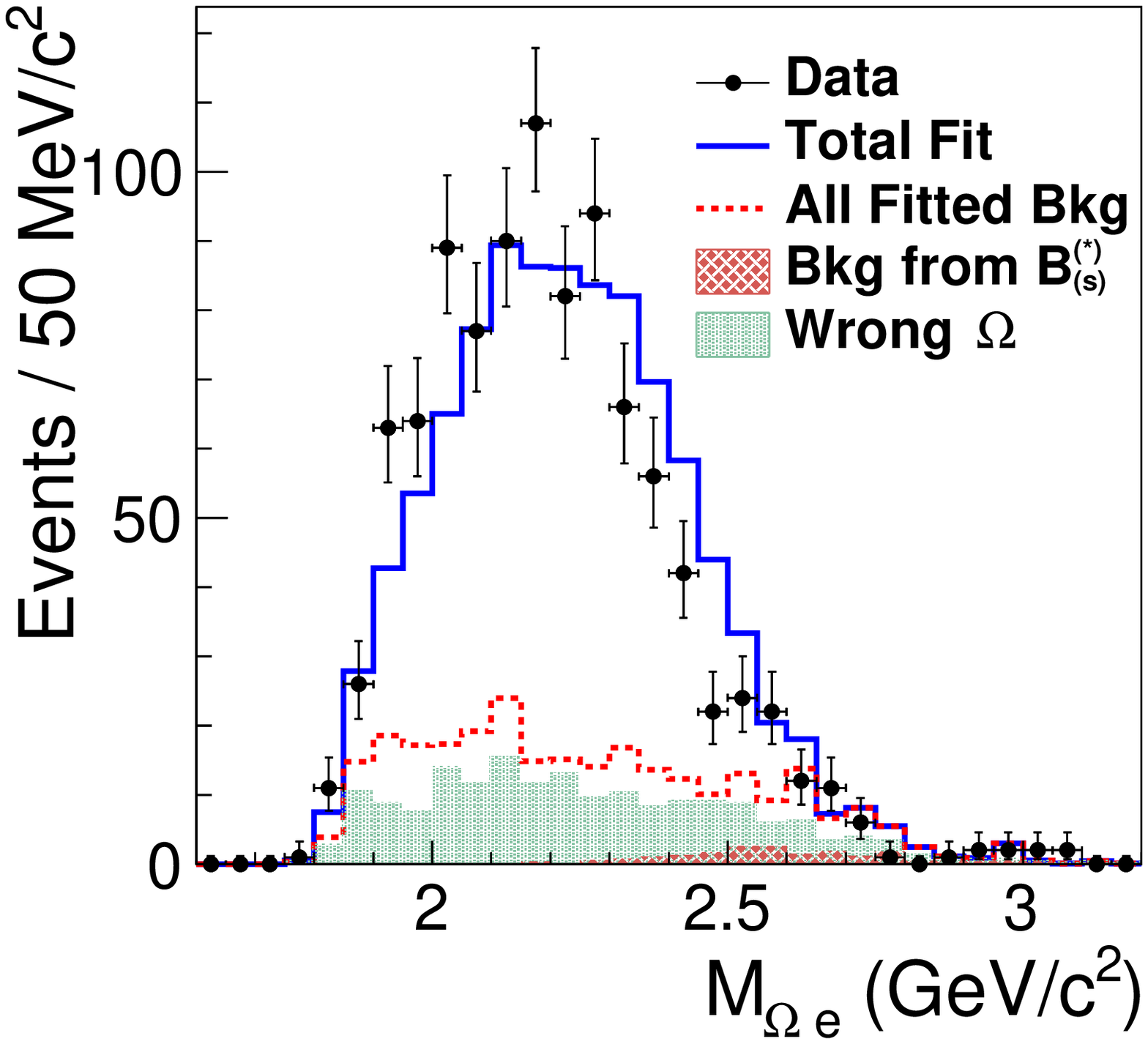}
		\includegraphics[width=5cm]{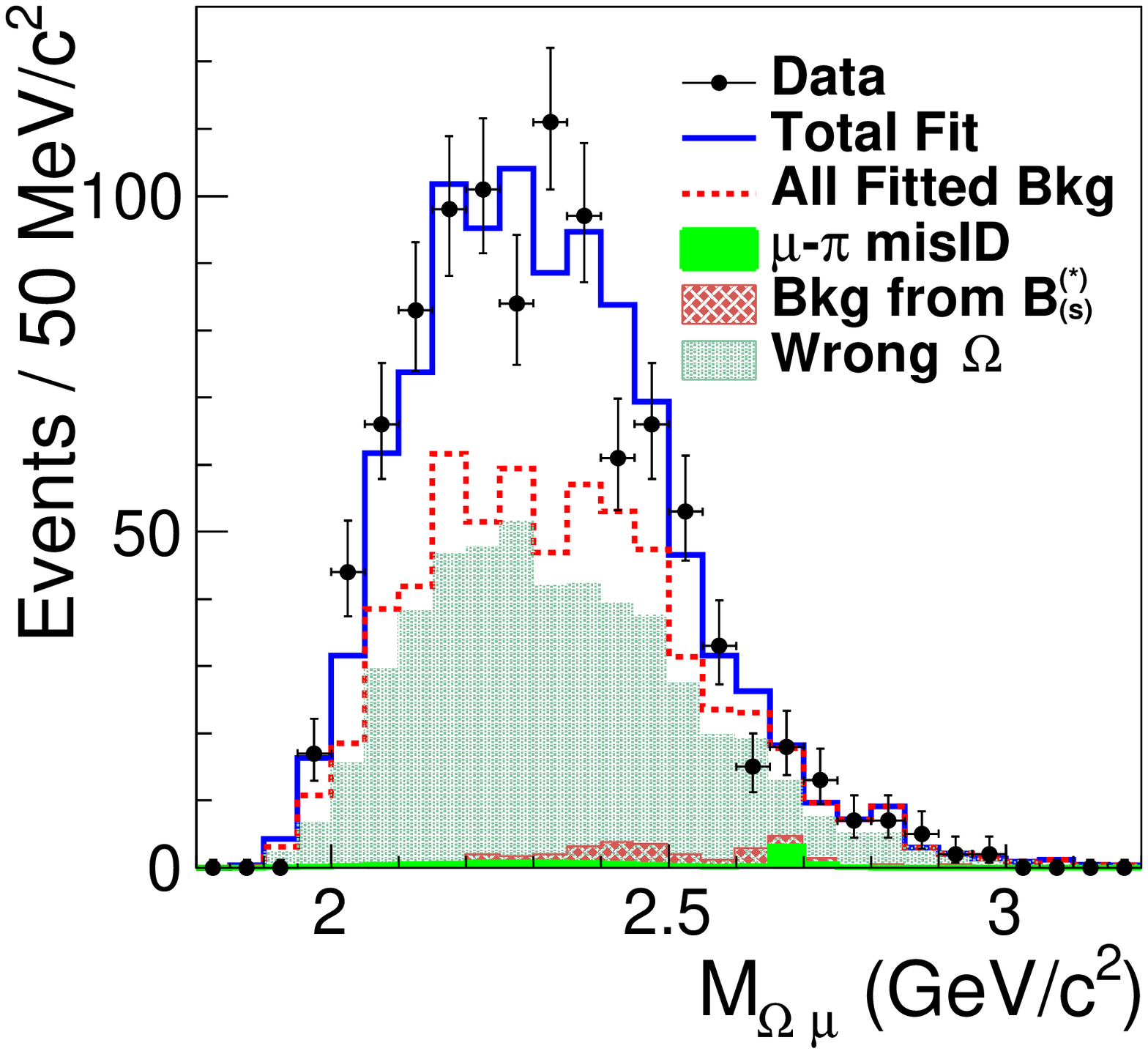}
		\put(-395,90){\bf (a)}
        \put(-255,90){\bf (b)}
		\put(-110,90){\bf (c)}
		\caption{ The fits to the (a) $M_{\Omega\pi}$, (b) $M_{\Omega e}$, and (c) $M_{\Omega \mu}$ distributions for the selected candidates from data. The dots with error bars represent the data, the solid lines are the best fits, and the dashed lines are the fitted total backgrounds. The blank areas between the red dashed lines and shaded histograms are from backgrounds with mis-selected $\ell^{+}$. The ``$\mu-\pi$ misID" in plot (c) means the background component from $\Omega_{c}^{0} \to \Omega^{-} \pi^{+}$$+$ $hadrons$ decays. The other fit components are illustrated by the legends.}\label{fit_br}
	\end{center}
\end{figure*}

For $\Omega_{c}^{0} \to \Omega^{-} e^{+} \nu_{e}$, the cosine of the opening angle
between $\Omega^{-}$ and $e^{+}$ in the lab frame is further required to be in the region (0.2, 0.95)
and the momentum of the $e^{+}$ in the center-of-mass frame is required to be in the region (0.35, 1.5) GeV/$c$.
For $\Omega_{c}^{0} \to \Omega^{-} \mu^{+} \nu_{\mu}$, the cosine of the opening angle
between $\Omega^{-}$ and $\mu^{+}$ in the lab frame is required to be larger than 0.35
and the momentum of the $\mu^{+}$ in the center-of-mass frame must be less than 1.6 GeV/$c$.

To suppress combinatorial backgrounds in each of the $\Omega e^{+} \nu_{e}$, $\Omega^{-} \mu^{+} \nu_{\mu}$, and $\Omega^{-} \pi^{+}$ modes, we require the scaled momentum $x_{p} = p^{*}_{\Omega X}/p^{*}_{\rm max} >0.5$, where $p^{*}_{\Omega X}$ is the momentum of the $\Omega X$ system in the center-of-mass frame (for $X = e,~\mu$ and $\pi$, respectively), and $p^{*}_{\rm max} \equiv \sqrt{{E_{\rm beam}}^{2} - {(m_{\Omega_{c}^{0}})^{2}}}$~\cite{Defc} ($E_{\rm beam}$ is the beam energy in the center-of-mass frame). This requirement removes all correct $\Omega X$ combinations from $\Omega_{c}^{0}$ produced in $B_{(s)}^{(*)}$ decays.

%%%%%%%%%%%%%%%%%%%%%%%%%%%%%%%%%%%%%
%%%%%%% semileptonic    %%%%%%%%%%%%%
%%%%%%%%%%%%%%%%%%%%%%%%%%%%%%%%%%%%%

After the above selections, the obtained $M_{\Omega\pi}$, $M_{\Omega e}$, and $M_{\Omega \mu}$
mass spectra from the data samples are shown in Fig.~\ref{fit_br}.
The $\Omega_{c}^{0}$ signals are extracted by binned maximum-likelihood fits to
these invariant mass spectra. In fitting the $M_{\Omega\pi}$ spectrum, the $\Omega_{c}^{0}$ signal shape is described by a double-Gaussian function with same mean value,
while the background shape is represented with a 1st-order polynomial, where all the parameters are floated. For $\Omega_{c}^{0}$ semileptonic decays, the signal shapes are taken directly from MC simulations. The background shapes from wrongly reconstructed $\Omega^{-}$ candidates are described by the $M_{\Omega^{-} \ell^{+}}$ distributions of $\Omega^{-}$ mass sidebands. The backgrounds from $e^+ e^- \to q\bar{q}$ due to mis-selected $\ell^{+}$ are represented by the $M_{\Omega^{-} \ell^{-}}$ distributions of $\Omega^{-}\ell^{-}$ events with their normalized $\Omega^{-}$ mass sidebands subtracted. The other backgrounds are from $e^+ e^- \to B_{(s)}^{(\ast)}\bar{B}_{(s)}^{(\ast)} + anything$ with $\Omega^{-}$ from one $B_{(s)}^{(\ast)}$ and $\ell^{+}$ from another $\bar{B}_{(s)}^{(\ast)}$, whose shapes are taken from simulated data. Background from $\Omega_{c}^{0} \to \Omega^{-} \pi^{0} \ell^{+} \nu_{\ell}$ decay is negligible since it violates isospin conservation and should be strongly suppressed. In fitting the $\Omega^{-} \mu^{+}$ mass spectrum, the $\mu-\pi$ misidentification background component stemming from $\Omega_{c}^{0} \to \Omega^{-} \pi^{+}$$+$ $hadrons$ events is added, with the relevant decay widths set to the PDG values~\cite{PDG}. In the fits to $M_{\Omega \ell}$ spectra above, the shapes of all fit components are fixed, and the yields are floated except for the backgrounds from wrongly reconstructed $\Omega^{-}$ candidates whose yields are normalized according to the $\Omega^{-}$ invariant mass distribution.
Figure~\ref{fit_br} shows the fitted results for $\Omega_{c}^{0}$ decays to (a) $\Omega^{-} \pi^{+}$, (b) $\Omega^{-} e^{+} \nu_{e}$, and (c) $\Omega^{-} \mu^{+} \nu_{\mu}$.
The fitted results together with the corresponding detection efficiencies are listed in Table~\ref{tab:br}. The efficiencies are computed on simulations and are then corrected to take into account data/MC discrepancies in the particle identifications (PID), where details will be explained in the section dedicated to the systematic uncertainty description. The significances of the $\Omega_{c}^{0} \to \Omega^{-}\ell^{+}\nu_{\ell}$ are both larger than 10$\sigma$. The significances are calculated using $\sqrt{-2\ln(\mathcal{L}_{0}/\mathcal{L}_{\rm max})}$, where $\mathcal{L}_{0}$ and $\mathcal{L}_{\rm max}$ are the likelihoods of the fits without and with a signal component, respectively.

\begin{table}[htbp]
	\caption{\label{tab:br} List of the fitted signal yields and the corresponding
detection efficiencies with the particle identification correction factors included. The last column gives the ratios of branching fractions $R = \BR(\Omega_{c}^{0} \to \Omega^{-}\ell^{+} \nu_{\ell})/\BR(\Omega_{c}^{0} \to \Omega^{-} \pi^{+})$. The branching fractions of $\Omega^{-} \to \Lambda K $ and $\Lambda \to p \pi^{-}$ are not included in the detection efficiencies. Quoted uncertainties are statistical only.}
		\renewcommand\arraystretch{1.2}
    	\begin{tabular}{lccc}
	\hline \hline
	channel &     signal yields &   detection efficiency   &    $R$   \\
	\hline
	$\Omega_{c}^{0} \to \Omega^{-} \pi^{+} $  & $865.3\pm 35.3$  &  $17.87\%$ &...\\
	$\Omega_{c}^{0} \to \Omega^{-} e^{+} \nu_{e}$ & $707.6 \pm 37.7 $& $7.40\%$  &$1.98\pm 0.13$\\
	$\Omega_{c}^{0} \to \Omega^{-} \mu^{+} \nu_{\mu}$  &$367.9\pm 31.4$ & $3.93\%$ &$1.94\pm 0.18$\\
	\hline \hline
\end{tabular}	

\end{table}

The $\Omega_{c}^{0}$ semileptonic decay branching fraction ratios are calculated using
$$
\frac{\BR(\Omega_{c}^{0} \to \Omega^{-} \ell^{+} \nu_{\ell})}{\BR(\Omega_{c}^{0} \to \Omega^{-} \pi^{+})}= \frac{N_{\Omega\ell}\cdot\varepsilon_{\Omega\pi}}{N_{\Omega\pi}\cdot\varepsilon_{\Omega\ell}},
$$
where $N$ and $\varepsilon$ are the fitted signal yields and detector efficiency of the corresponding $\Omega_{c}^{0}$ decay, respectively. The calculated results are listed in Table~\ref{tab:br}. Similarly, we also obtain $\BR(\Omega_{c}^{0} \to \Omega^{-} e^{+} \nu_{e})/\BR(\Omega_{c}^{0} \to \Omega^{-} \mu^{+} \nu_{\mu}) = 1.02 \pm 0.10$. Here, the uncertainties are statistical only.

%%%%%%%%%%%%%%%%%%%%%%%%%%%%%%%%%%%%%
%%%%%%  systematic error  %%%%%%%%%%%
%%%%%%%%%%%%%%%%%%%%%%%%%%%%%%%%%%%%%

Several sources of systematic uncertainties contribute to the measurement of the branching fraction ratios. Using $D^{*+} \to D^{0}
\pi^{+}$, $D^{0} \to K^{-} \pi^{+}$, and $J/\psi \to \ell \ell$ control samples, the efficiency ratios between data and MC simulations are $(95.4 \pm 0.9)\%$, $(98.2 \pm 0.9)\%$, and $(98.7 \pm 0.6)\%$ for pion, electron, and muon, respectively. The central values of the ratios are taken as efficiency correction factors and the relative errors are taken as systematic uncertainties, written as $\sigma_{\rm PID}$ in Table~\ref{tab:err}. The systematic uncertainties associated with tracking efficiency and $\Omega^{-}$ selection approximately cancel in the branching fraction ratio measurements so that the uncertainties on those are negligible.
We estimate the systematic uncertainties associated with the fitting procedures ($\sigma_{\rm fit}$) for $\Omega_{c}^{0} \to \Omega^{-} \ell^{+} \nu_{\ell}$ and
$\Omega_{c}^{0}\to \Omega^{-} \pi^{+}$ separately.
For $\Omega_{c}^{0} \to \Omega^{-} \ell^{+} \nu_{\ell}$ decays, we change the bin width of the $M_{\Omega \ell}$ spectra by $\pm$5 MeV/$c^{2}$,
change the $\Omega^{-}$ mass sidebands from four to three times that of the signal region,  %add the background component from $\Xi_{c} \to \Xi \pi^{+} \pi^{-} \ell^{+} \nu_{\ell}$ with its shape taken from MC simulation and yields floated},
and take the relative differences of the fitted signal yields as
$\sigma_{\rm fit}$: these are $1.0\%$ for the electron mode and muon mode.
For $\Omega_{c}^{0}\to \Omega^{-} \pi^{+}$, we estimate $\sigma_{\rm fit}$ by changing the range of the fit and the order of the background polynomial, and take the relative difference of the signal yields, $0.4\%$, as the systematic uncertainty.
For $\Omega_c^{0} \to \Omega^{-} \pi^{+}$, the $x_{p}$ distribution is corrected with efficiencies bin by bin, and is fitted with Peterson's fragmentation function $1/(x_{p} \cdot (1-\frac{1}{x_{p}}-\frac{\epsilon_{p}}{1-x_{p}})^{2})$~\cite{FF_theory}. The signal MC samples of all three decay modes are generated with the fitted Peterson's fragmentation function, and the relative difference of the detection efficiencies obtained by changing the fitted $\epsilon_{p}$ by $\pm1\sigma$ are taken as the systematic uncertainty ($\sigma_{x_p}$), which are 0.5\%, 0.5\%, and 2.1\% for electron, muon, and pion mode, respectively.
For semileptonic decays, to conservatively estimate the uncertainties due to possible imperfect modeling by the {\sc pythia} matrix element model, the signal MC samples of $\Omega_{c}^{0}\to\Omega^{-}\ell^{+}\nu_{\ell}$ decays are simulated with phase space model. The changes in measured $N_{\Omega \ell}/\varepsilon_{\Omega \ell}$ are taken as the uncertainties related to the MC model ($\sigma_{\rm MC}$), which are 2.6\% and 3.0\% for the electron and muon mode, respectively. The relative changes of the $N_{\Omega \ell}/\varepsilon_{\Omega \ell}$ by fitting the $M_{\Omega\ell}$ spectra without the background component from $B_{(s)}^{(\ast)}$ decays are taken as the uncertainties due to $B_{(s)}^{(\ast)}$ decay ($\sigma_{B_{(s)}^{(\ast)}}$), which are 1.4\% and 2.8\% for the electron and muon mode, respectively.
The corresponding systematic uncertainties are summed in quadrature to yield the total systematic uncertainty ($\sigma_{N/\varepsilon}^{\rm tot}$) for each $\Omega_{c}^{0}$ decay mode, which yields 2.3\%, 3.3\%, and 4.3\% for the pion, electron, and muon mode, respectively.
The relative systematic uncertainties described above are summarized in Table~\ref{tab:err}.
\begin{table}[htbp]
	\caption{\label{tab:err} The relative systematic uncertainties on $N_{\Omega X}/\varepsilon_{\Omega X}$, where the common systematic uncertainties in all the decay channels are not included (\%). }
	\tabcolsep=5pt
	\renewcommand\arraystretch{1.2}
	\begin{tabular}{lcccccc}
		\hline \hline
		channel &     $\sigma_{\rm PID}$ &  $\sigma_{\rm fit}$   & $\sigma_{x_p}$& $\sigma_{\rm MC}$ &$\sigma_{B_{(s)}^{(\ast)}}$ & $\sigma_{N/\varepsilon}^{\rm tot}$\\
		\hline
		$\Omega_{c}^{0} \to \Omega^{-} \pi^{+} $  & $0.9$  &  $0.4$ &$2.1$& ... &...&2.3\\
		$\Omega_{c}^{0} \to \Omega^{-} e^{+} \nu_{e}$ & $0.9 $& $1.0$  &0.5& $2.6$ &$1.4$ &3.3\\
		$\Omega_{c}^{0} \to \Omega^{-} \mu^{+} \nu_{\mu}$  & $0.6 $& $1.0$  &0.5& $3.0$ &$2.8$ &$4.3$\\
		\hline \hline
	\end{tabular}	
	
\end{table}

 The final systematic uncertainty of the branching fraction ratio is the sum of the corresponding two $\sigma_{N/\varepsilon}^{\rm tot}$ in quadrature, which yields 4.0\%  and 4.9\% for $\BR(\Omega_{c}^{0} \to \Omega^{-}\ell^{+} \nu_{\ell})/\BR(\Omega_{c}^{0} \to \Omega^{-} \pi^{+})$, with $\ell^{+} = e^{+}$ and $\mu^{+}$, respectively. The total systematic uncertainty on $\BR(\Omega_{c}^{0} \to \Omega^{-}e^{+} \nu_{e})/\BR(\Omega_{c}^{0} \to \Omega^{-} \mu^{+} \nu_{\mu})$ is 2.3\% with the $\sigma_{B_{(s)}^{(\ast)}}$, $\sigma_{x_p}$, and $\sigma_{\rm MC}$ positively correlated.

%%%%%%%%%%%%%%%%%%%%%%%%%%%%%%%%%%%%%
%%%%%%%%%%% summary  %%%%%%%%%%%%%%%%
%%%%%%%%%%%%%%%%%%%%%%%%%%%%%%%%%%%%%

According to the analysis above, the branching fraction ratios $\BR(\Omega_{c}^{0} \to \Omega^{-} e^{+} \nu_{e})/\BR(\Omega_{c}^{0} \to \Omega^{-} \pip)$ and $\BR(\Omega_{c}^{0} \to \Omega^{-} \mu^{+} \nu_{\mu})/\BR(\Omega_{c}^{0} \to \Omega^{-} \pip)$ are measured to be $1.98 \pm 0.13 \pm 0.08$ and $ 1.94 \pm 0.18 \pm 0.10$, respectively.  The ratio $\BR(\Omega_{c}^{0} \to \Omega^{-} e^{+} \nu_{e})/\BR(\Omega_{c}^{0} \to \Omega^{-} \pip)$ is consistent with the previously measured value $ 2.4 \pm 1.2$ by the CLEO collaboration~\cite{cleo_omgc} with greatly improved precision. The Ratio of $\BR(\Omega_{c}^{0} \to \Omega^{-} e^{+} \nu_{e})/\BR(\Omega_{c}^{0} \to \Omega^{-} \mu^{+} \nu_{\mu})$ is measured to be $1.02 \pm 0.10 \pm 0.02$, which is consistent with the expected LFU value $1.03\pm0.06$~\cite{qian}.	Here, the first and second uncertainties are statistical and systematic, respectively.

In summary, based on data samples of 89.5, 711  and 121.1 fb$^{-1}$ collected with the Belle detector at the center-of-mass energies of 10.52, 10.58, and 10.86 GeV, respectively, we measured branching fraction ratios of $\BR(\Omega_{c}^{0} \to \Omega^{-} \ell^{+} \nu_{\ell})/\BR(\Omega_{c}^{0} \to \Omega^{-} \pi^{+})$ and $\BR(\Omega_{c}^{0} \to \Omega^{-} e^{+} \nu_{e})/\BR(\Omega_{c}^{0} \to \Omega^{-} \mu^{+} \nu_{\mu})$. The $\Omega_{c}^{0} \to \Omega^{-} \mu^{+} \nu_{\mu}$ decay is seen for the first time.  Our measured $\BR(\Omega_{c}^{0} \to \Omega^{-} \ell^{+} \nu_{\ell})/\BR(\Omega_{c}^{0} \to \Omega^{-} \pip)$ are larger than those from the predictions of the light-front quark models~\cite{epjcOmgc_sl,qian}, and $\BR(\Omega_{c}^{0} \to \Omega^{-} e^{+} \nu_{e})/\BR(\Omega_{c}^{0} \to \Omega^{-} \mu^{+} \nu_{\mu})$ agrees with the expectation of LFU. The semileptonic branching fraction ratio $\BR(\Omega_{c}^{0} \to \Omega^{-} \ell^{+} \nu_{\ell})/\BR(\Omega_{c}^{0} \to \Omega^{-} \pi^{+})$  is an important input used to constrain parameters of  phenomenological models~\cite{epjcOmgc_sl,qian,prcOmgc_sl,plbOmgc_sl} and the ongoing lattice QCD calculations of heavy flavor baryon decays. Once measurement of the absolute branching fraction of ${\cal B}(\Omega_{c}^{0} \to \Omega^{-}\pi^{+})$ become available in the near future, the results presented in this Letter will lead to the value of $\BR(\Omega_{c}^{0} \to \Omega^{-} \ell^{+} \nu_{\ell})$ which can be compared with more theoretical expectations and with those of other semileptonic decays of charmed baryons.

%%%%%%%%%%%%%%%%%%%%%%%%%%%%%%%%%%%%%
%%%%%%%   Acknowledgments   %%%%%%%%%
%%%%%%%%%%%%%%%%%%%%%%%%%%%%%%%%%%%%%
Y. B. Li acknowledges the support from China
Postdoctoral Science Foundation (2020TQ0079).
We thank the KEKB group for excellent operation of the
accelerator; the KEK cryogenics group for efficient solenoid
operations; and the KEK computer group, the NII, and
PNNL/EMSL for valuable computing and SINET5 network support.
We acknowledge support from MEXT, JSPS and Nagoya's TLPRC (Japan);	
ARC (Australia); FWF (Austria); the
National Natural Science Foundation of China under
Contracts No. 11575017, No. 11761141009, No. 11975076, No. 12042509, No. 12135005, No. 12161141008;
MSMT (Czechia);
ERC Advanced Grant 884719 and Starting Grant 947006 (European Union);
CZF, DFG, EXC153, and VS (Germany);
DAE (Project Identification No. RTI 4002) and DST (India); INFN (Italy);
MOE, MSIP, NRF, RSRI, FLRFAS project, GSDC of KISTI and KREONET/GLORIAD (Korea);
MNiSW and NCN (Poland); MSHE, Agreement No. 14.W03.31.0026, and HSE UBRC (Russia);
University of Tabuk (Saudi Arabia); ARRS Grants J1-9124 and P1-0135 (Slovenia);
IKERBASQUE (Spain);
SNSF (Switzerland); MOE and MOST (Taiwan); and DOE and NSF (USA).


\begin{thebibliography}{99}
\bibitem{Defc} We used units in which the speed of light is $c = 1$.
\end{thebibliography}
\end{document}